\documentclass[letters, fleqn,usenatbib, useAMS]{mnras}
\usepackage{graphicx}	% Including figure files
\usepackage{amsmath}	% Advanced maths commands
\usepackage{amssymb}	% Extra maths symbols
\usepackage{multicol}        % Multi-column entries in tables
\usepackage{bm}		% Bold maths symbols, including upright Greek
\usepackage{listings}
\usepackage{alltt}
\usepackage{booktabs}
\usepackage{lipsum}
\usepackage{color}
\usepackage{txfonts}
\usepackage{times}

\title[A Malin 1 analogue in IllustrisTNG]{Formation of a Malin 1 analogue in IllustrisTNG by stimulated accretion}

\author[Q. Zhu et al.]
{Qirong Zhu$^{1,2}$\thanks{E-mail: qxz125@psu.edu}, 
Dandan Xu$^3$,
Massimo Gaspari$^4$\thanks{Einstein and Spitzer Fellow},
Vicente Rodriguez-Gomez$^5$, \newauthor
Dylan Nelson$^{6}$,
Mark Vogelsberger$^{7}$\thanks{Alfred P.~Sloan Fellow},
Paul Torrey$^{7}$\thanks{Hubble Fellow}, 
Annalisa Pillepich$^{8}$,  \newauthor
Jolanta Zjupa$^{3, 9}$, 
Rainer Weinberger$^{3}$, 
Federico Marinacci$^{7}$, 
R\"udiger~Pakmor$^3$,  \newauthor
Shy Genel$^{10, 11}$, 
Yuexing Li$^{1}$,
Volker Springel$^{3, 6,12}$,
and Lars Hernquist$^{2}$\\
\vspace{0.15cm}\\
\parbox{\textwidth}
{\small $^1$ Department of Astronomy \& Astrophysics; Institute for Cosmology and Gravity, The Pennsylvania State University, PA 16802, USA\\
$^2$ Harvard-Smithsonian Center for Astrophysics, Harvard University, 60 Garden Street, Cambridge, MA 02138, USA\\
$^3$ Heidelberger Institut f\"ur Theoretische Studien, Schloss-Wolfsbrunnenweg 35, 69118 Heidelberg, Germany\\
$^4$ Department of Astrophysical Sciences, Princeton University, 4 Ivy Lane, Princeton, NJ 08544-1001 USA\\
$^5$ Department of Physics \& Astronomy, Johns Hopkins University, 3400 N. Charles Street, Baltimore, MD 21218, USA\\
$^6$ Max-Planck-Institut f\"ur Astrophysik, Karl-Schwarzschild-Str. 1, D-85748, Garching, Germany\\
$^7$ Department of Physics, Kavli Institute for Astrophysics and Space Research, MIT, Cambridge, MA 02139, USA\\
$^8$ Max-Planck-Institut f\"ur Astronomie, K\"onigstuhl 17, 69117 Heidelberg, Germany\\
$^9$ Institut f\"ur Theoretische Physik, Philosophenweg 16, 69120 Heidelberg, Germany\\
$^{10}$ Center for Computational Astrophysics, Flatiron Institute, 162 Fifth Avenue, New York, NY 10010, USA\\
$^{11}$ Columbia Astrophysics Laboratory, Columbia University, 550 West 120th Street, New York, NY 10027, USA\\
$^{12}$ Zentrum f\"ur Astronomie der Universit\"at Heidelberg, ARI, M\"onchhofstrasse 12-14, 69120 Heidelberg, Germany\\
}}

\begin{document}
%\date{Accepted XXX. Received YYY; in original form ZZZ}
\pubyear{2018}
\pagerange{\pageref{firstpage}--\pageref{lastpage}} 

\maketitle

\label{firstpage}

\begin{abstract}
The galaxy Malin 1 contains the largest stellar disk known but 
the formation mechanism of this structure has been elusive.
In this paper, we report a Malin 1 analogue in the 100 
Mpc IllustrisTNG simulation and describe its formation history.  
At redshift zero, this massive galaxy, having a maximum circular 
velocity $V_{\rm max}$ of 430 ${\rm km\ s^{-1}}$, contains a
100 kpc gas/stellar disk with morphology similar to Malin 1. 
The simulated galaxy reproduces well many observed features of 
Malin 1's vast disk, including its stellar ages, metallicities, and gas 
rotation curve. We trace the extended disk back in time and find that
a large fraction of the cold gas at redshift zero originated from the 
cooling of hot halo gas, triggered by the merger of a pair of intruding 
galaxies. Our finding provides a novel way to form large galaxy disks 
as extreme as Malin 1 within the current galaxy formation framework. 
\end{abstract}


\section{Introduction} 
\label{sec:intro}
Malin 1, discovered by \cite{Bothun1987}, has one of the largest stellar 
disks known. With a diameter of at least 200 kpc,  its angular size 
on the sky is wider than one arcmin even though Malin 1 is a quite 
distant object ($z\sim0.09$). Malin 1 is a prime example of  
giant low surface brightness (GLSB) galaxies, characterized by their 
extended low surface brightness disk and high gas content 
\citep[]{Sprayberry1995}. Other examples of extreme GLSB 
galaxies exist~\citep[e.g. UGC1382,][]{Hagen2016} with properties 
similar to Malin 1. Without the extended outer disks, Malin 1 and 
UGC1382 would otherwise appear similar to normal massive 
galaxies \citep{Barth2007, Hagen2016}.

Observations of the Malin 1 disk are still limited due to its low 
surface brightness. Recent observations by \cite{Boissier2016} 
show spectacular spiral structure made of relatively young 
($\sim0.1$--3 Gyr) stars with metallicities between $0.1-1$ 
$Z_{\odot}$ with no apparent stellar age gradient across the 
disk. A more surprising feature is the large mass with $V_{\rm max}$ 
above $400\ {\rm km\ s^{-1}}$ if the (poorly constrained) inclination 
is low~\citep[consistent with, e.g.,][]{Galaz2015}.

It remains a great challenge to understand the physical process 
behind a disk that is at least five times larger than the 
Milky Way. This single object, being an extreme outlier, is a 
severe test for the current theory of galaxy formation and modified 
gravity \citep{Lelli2010,Boissier2016}. 
In this \textit{Letter}, we report a Malin 1 analogue from the 
IllustrisTNG simulations. At $z=0$, this galaxy 
has an extended gas/stellar 
disk with size/morphology similar to Malin 1. 
We then trace the extended disk back in time and find that a large fraction 
of the neutral gas came from the cooling of hot halo gas, triggered by 
a merger event involving \textit{a pair of intruding galaxies}, a process we
term `stimulated accretion'. In what follows, 
we first describe our methods and present our main results in 
Section~\ref{sec:methodsandresults}. We discuss our findings and
present observational tests in Section~\ref{sec:discussions} and conclude 
in Section~\ref{sec:conclusions}. 

\begin{figure}
\begin{center}
\vspace{-0.2cm}
\centerline{\vbox{\hbox{
\includegraphics[width=0.24\textwidth]{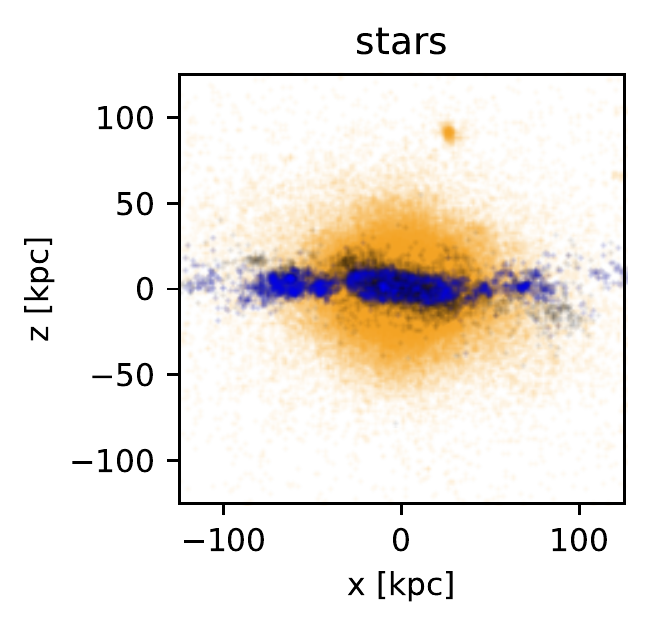}
\includegraphics[width=0.24\textwidth]{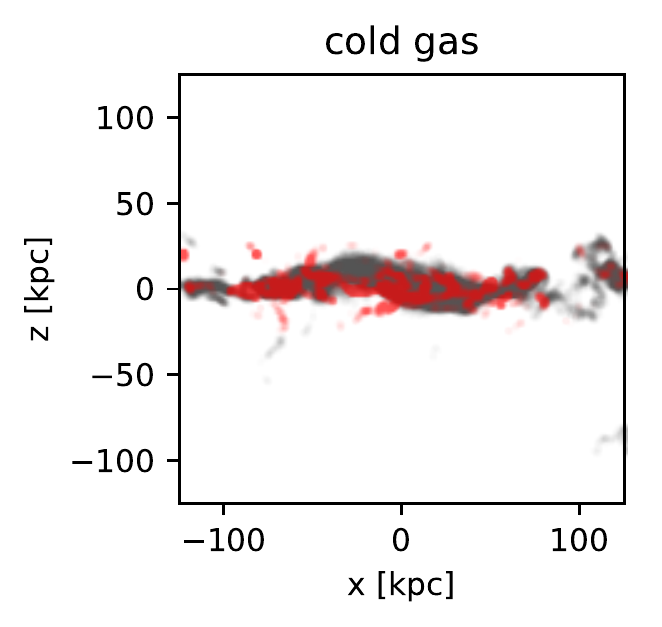}}}}
\vspace{-0.2cm}
\centerline{\vbox{\hbox{
\includegraphics[width=0.24\textwidth]{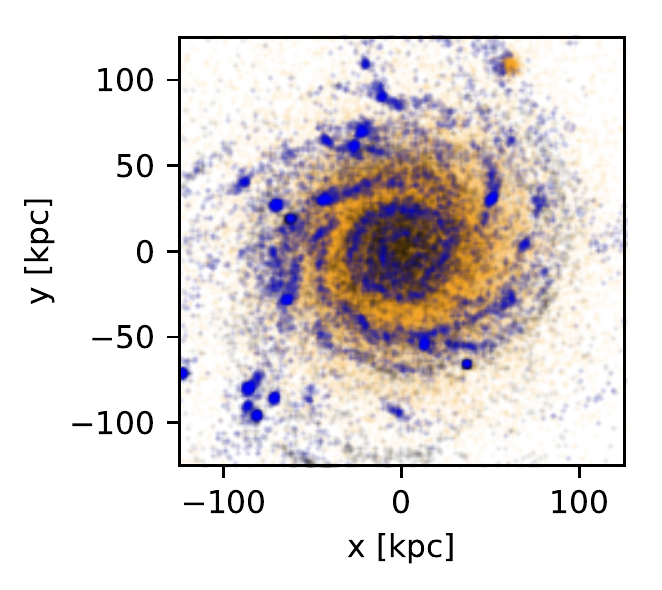}
\includegraphics[width=0.24\textwidth]{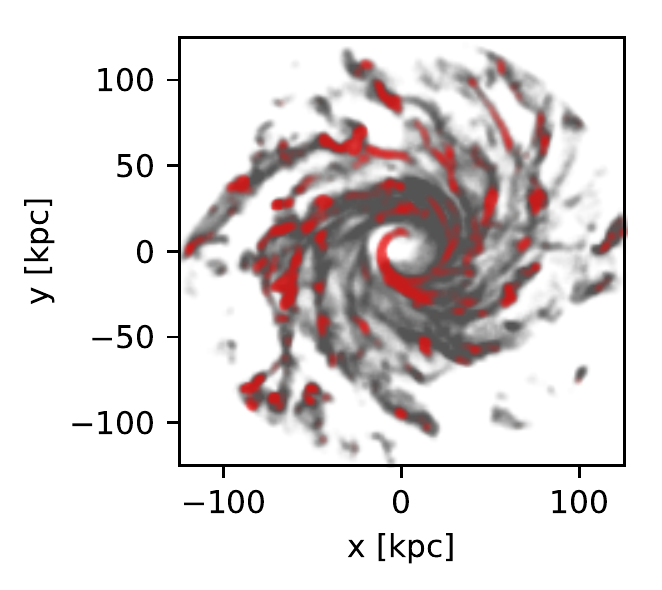}}}}
\vspace{-0.2cm}
\centerline{\vbox{\hbox{
\includegraphics[width=0.24\textwidth]{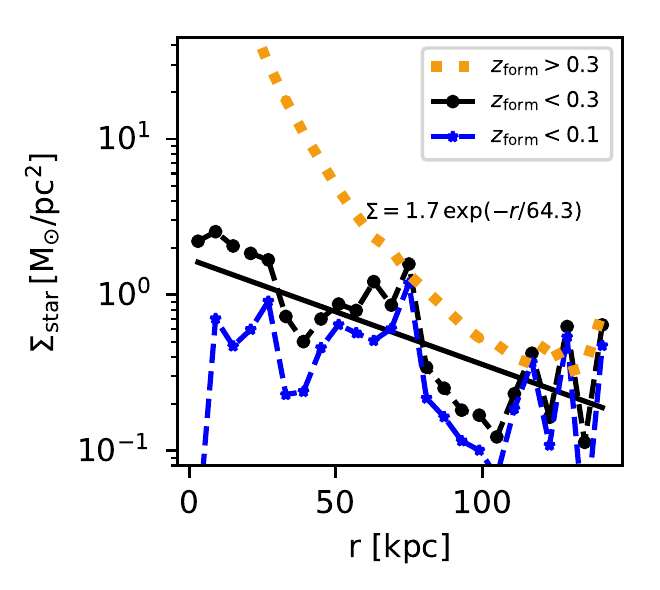}
\includegraphics[width=0.24\textwidth]{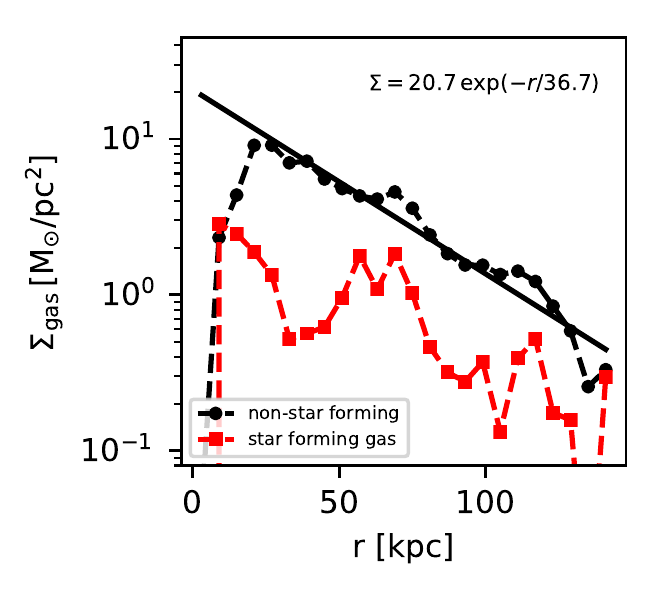}}}}
\end{center}
\vspace{-0.5cm}
\caption{\label{fig:galaxy_images}
Edge-on (top row) and face-on (middle row) views of the stellar (left) and gas (right) disk. 
Newly formed stars are shown in blue ($z_{\rm form} < 0.1$)  and 
gray ($z_{\rm form} < 0.3$) while old stars in orange.
The stellar surface density with $z_{\rm form} < 0.3$ follows an 
exponential profile with a scale length of 64 kpc. 
Outside of 100 kpc, the newly formed stellar disk
is comparable to the existing stellar mass (the dotted line) while 
the luminosity will be dominated by the young population. 
In the right column, we show the cold gas in gray
and star forming-gas in red. The surface density profile of the cold gas follows 
an exponential profile with a scale length of 37 kpc.}
\end{figure}

\vspace{-0.15in} 
\section{Methods and Results}
\label{sec:methodsandresults}

\begin{figure}
\begin{center}
\includegraphics[width=0.8\linewidth]{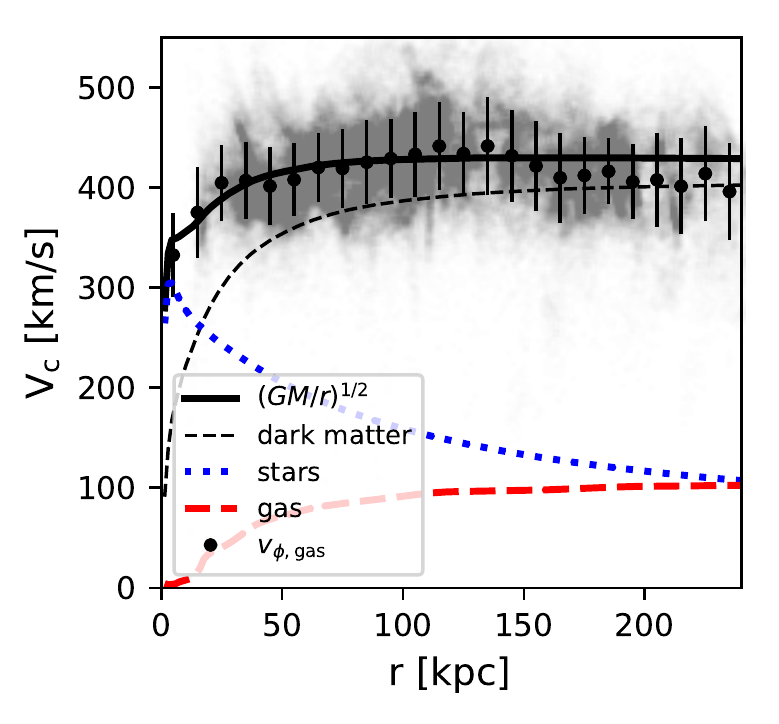}\\
\vspace{-0.2cm}
\caption{Circular velocity curve from the mass distribution as a function of 
radius. The tangential velocity of gas cells is overplotted with the gray scatter 
points. The extended gas disk at $z = 0$ is rotating close to the 
circular velocity curve, and is therefore rotationally supported. 
Within the inner 20 kpc, the stellar mass dominates the mass budget. 
}
\end{center}
\label{fig:rotation_curve}
\end{figure}

We identified this galaxy during our search for GLSB 
galaxies (Zhu et. al, in prep) in the IllustrisTNG simulation suite 
\citep{TNG5, TNG3, TNG1, TNG4, TNG2}. The IllustrisTNG 
simulations build on the original Illustris project through a series of numerical 
and physical model improvements \citep{Weinbergers2017, Pillepich2018} 
over the original Illustris model \citep{Vogelsberger2014b, Vogelsberger2014, 
Genel2014, Sijacki2015}. In this work, we use the TNG100 simulation which involves a
box of side length $\sim$ 100 Mpc as the original Illustris 
simulation.  The baryon mass resolution is $1.4 \times 10^6 M_{\odot}$ 
with a fixed softening length for stars of 0.7 kpc at z=0 and an adaptive 
softening for gas cells with a minimum of 0.185 comoving kpc. Throughout the text, 
cold gas refers to non-star forming gas with $\log(T{\rm /K})<4.2$ (to be
compared with neutral hydrogen), while hot gas refers to gas with 
$\log(T{\rm /K}) > 6$. Additionally, `star-forming gas' is used 
when the gas density exceeds the threshold for star formation, 
$n_{\rm sfr} = 0.13\ {\rm cm}^{-3}$.

To identify GLSB galaxies, we use two-component (bulge+disk) 
surface photometry fitting \citep{Xu2017} and create a candidate list. 
Through visual inspection, we identified one object 
(\textbf{Central}, hereafter), bearing a resemblance 
to Malin 1 in the stellar and gaseous morphologies. 
The {\small SUBFIND} ID of the Central at $z = 0$ is 252245. 
Not only is the size of the disk in this galaxy the largest in the list, 
it has the largest cold gas mass in the entire galaxy catalogue.

Fig.~\ref{fig:galaxy_images} shows the stellar 
and gas disks from edge-on and face-on views. 
Without the stars formed later than $z=0.3$, 
the Central is just a massive elliptical galaxy with a 
spherical stellar halo. The recently formed stars, on the other hand, 
are located in an extended disk with many apparent 
stellar ``knots". These knots are groups of newly formed stars from 
gas-rich clumps in the disk without associated dark matter haloes.
Most of the stars in the disk are formed within the last 1.5 Gyr 
($z_{\rm form}=0.1$, blue dots in Fig.~\ref{fig:galaxy_images}). 
Also, there is not any clear stellar age gradient across the disk. 
Many of the features including the stellar ``knots", young stellar population, 
spiral structure, and lack of age gradient agree well with the observations 
of Malin 1~\citep{Boissier2016}.

The gas distribution is in an extended and well-organized 
disk with a radius larger than 100 kpc. In the right panels of 
Fig.~\ref{fig:galaxy_images}, we show the cold gas 
in gray and star-forming gas in red.  An exponential profile
well describes the projected surface density profile of the gas disk, 
with a scale length of 37 kpc. Moreover, the 
surface density of the cold gas is one order of magnitude larger than 
the surface density of stars, highlighting the galaxy's gas-rich nature.

Figure~\ref{fig:rotation_curve} shows the circular velocity curves 
from the total mass distribution $V_c=\sqrt{GM(r)/r}$ as well as the 
contribution from each component. The tangential velocity of the gas 
cells in the disk plane is shown with the gray scatter points. The gas disk
rotates close to the circular velocity curve determined by the mass
distribution. The gas rotation curve stays flat up to 200 kpc, which 
qualitatively agrees with \cite{Lelli2010}.  Moreover, this galaxy has a 
$V_{\rm max}$ of $430\ {\rm km\ s^{-1}}$ which is in excellent agreement 
with the values derived by \cite{Boissier2016}.

\begin{figure}
\begin{center}
\includegraphics[width=0.8\linewidth]{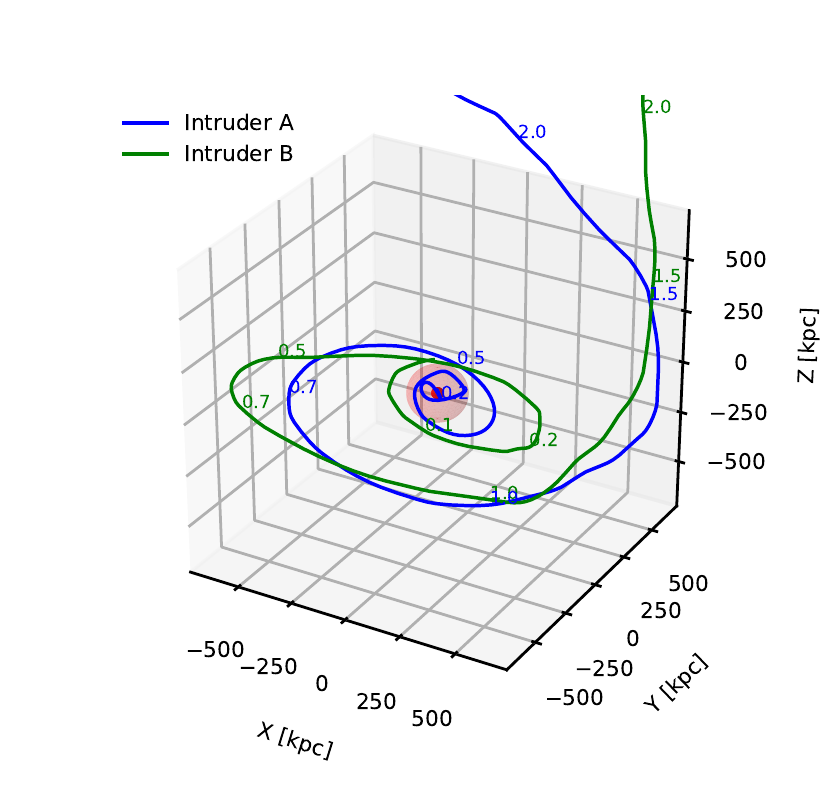}\\
\includegraphics[width=0.8\linewidth]{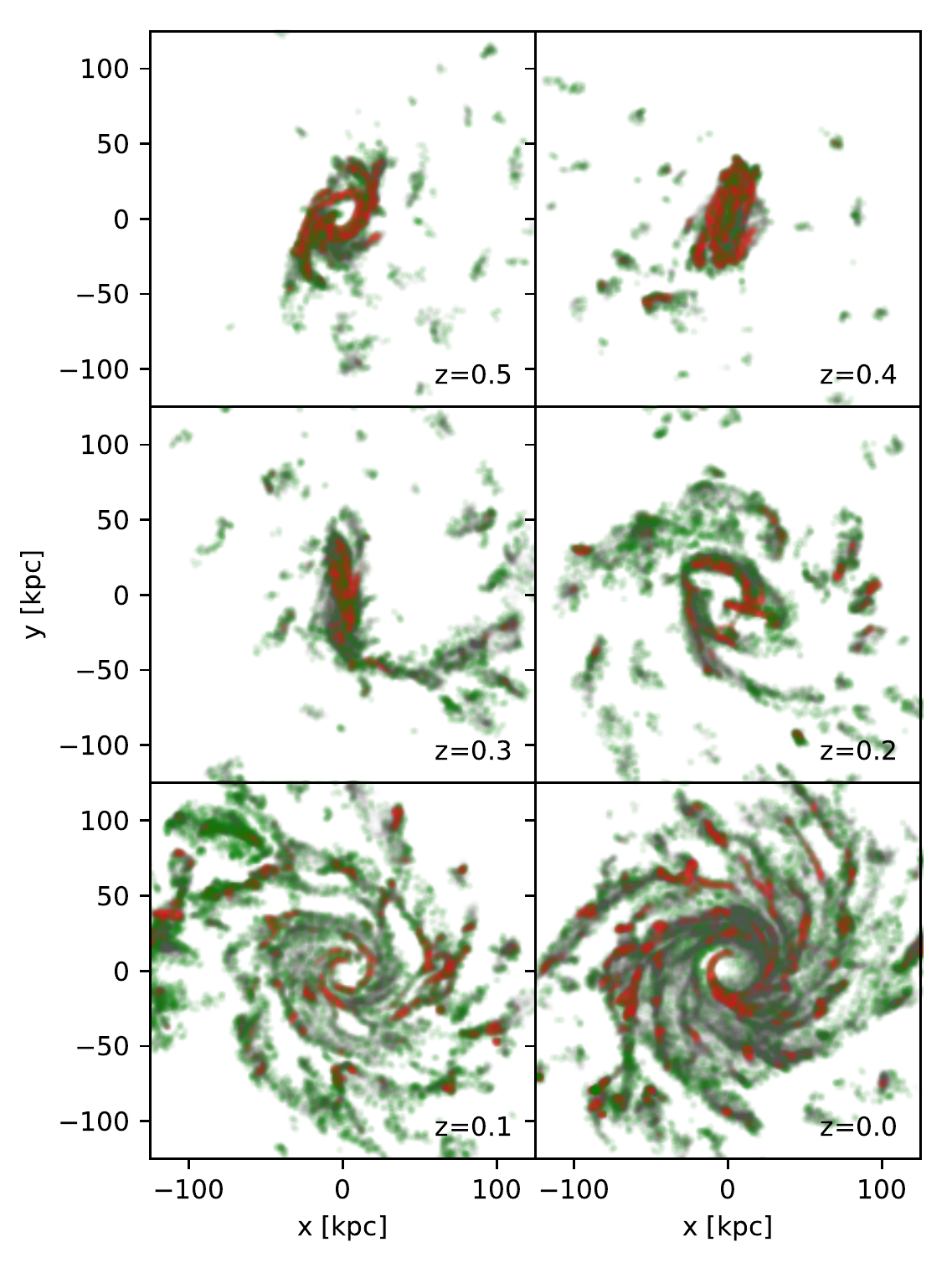}\\
\caption{\label{fig:merger}
\textit{Top half}: Orbits of the galaxy pair, Intruder A and B,  into the Central
from $z = 2$ to 0. The central red sphere has a radius of 125 kpc, the size of
the final gas disk. We also label several redshifts along with galaxy trajectories
to guide the readers.
\textit{Lower half}: The formation of a very extended gas 
disk from $z = 0.3$ to $z = 0$. In addition to cold gas (gray)
and star-forming gas (red), we show the gas cells 
with temperature between $10^5$ and $10^6$ K in green.
Intruder A, with a peak $V_{\rm max}=350\ {\rm km\ s^{-1}}$, sank
into the central potential faster than Intruder B, with a peak 
$V_{\rm max}=120\  {\rm km\ s^{-1}}$. While Intruder A is gas poor, 
there is much cold gas associated with Intruder B.
The leading gas arm of Intruder B follows the trajectory of Intruder A, 
seen as the one arm structure at $z = 0.3$, 
and triggers efficient cooling of hot halo gas, which is confirmed 
with the location of gas with temperature between $10^5$ and $10^6$~K. }
\end{center}
\end{figure}

%How did this extended gas disk form? 
To shed light on the formation mechanism of such an extended stellar/gas disk,
we use a merger tree \citep{RodriguezGomez2015} to trace back its 
formation history. Interestingly, we find the disk appeared only after 
$z = 0.3$, when a merger event occurred. The merger event involves
a pair of in-falling galaxies with peak circular velocities of 
350 and 120 ${\rm km\ s^{-1}}$ respectively. The top half of Fig.~\ref{fig:merger}
shows the orbits of the two galaxies, which were orbiting around each other 
between $z=1.5$ and 0.8 while inspiralling into the Central.
The more massive galaxy (\textbf{Intruder A})
sinks more rapidly in the potential well after $z = 0.5$
due to stronger dynamical friction
and completely merges with the Central at $z  = 0.16$. Meanwhile, 
this intruder also completely disrupts the original gas disk of 
the Central (was visible as an edge-on disk in the center
of the gas image at $z = 0.3$). Additionally, very little gas is directly
delivered by Intruder A, which has lost most of its gas mass since 
$z = 1$ when it was still fairly distant (625 comoving $\rm{kpc}$) 
from the Central.

The less massive intruder (\textbf{Intruder B}), on the other hand, 
is quite gas-rich. The interaction between the galaxies in the pair has 
already liberated a great deal of gas from this object to form an 
elongated gas trail started from $z = 0.4$. 
Part of the leading arm of the cold gas closely follows 
Intruder A. By $z = 0.3$, the leading arm has 
already penetrated into 100 kpc from the Central. 
Since then an extended rotating gas disk, as shown in 
the lower half of Fig.~\ref{fig:merger}, forms and builds up.
By $z = 0$, Intruder B has not been completely disrupted yet. 
To see where the cold gas came from,
we also color the gas cells with temperature 
between $10^5$ and $10^6$~K in green. The location of gas at 
such a transitional stage closely follows the cold gas in the leading arm, 
which suggests that efficient cooling may have taken place at the 
interface between the cold gas and the hot halo, which we will show next.

\begin{figure}
\begin{center}
\includegraphics[width=0.8\linewidth]{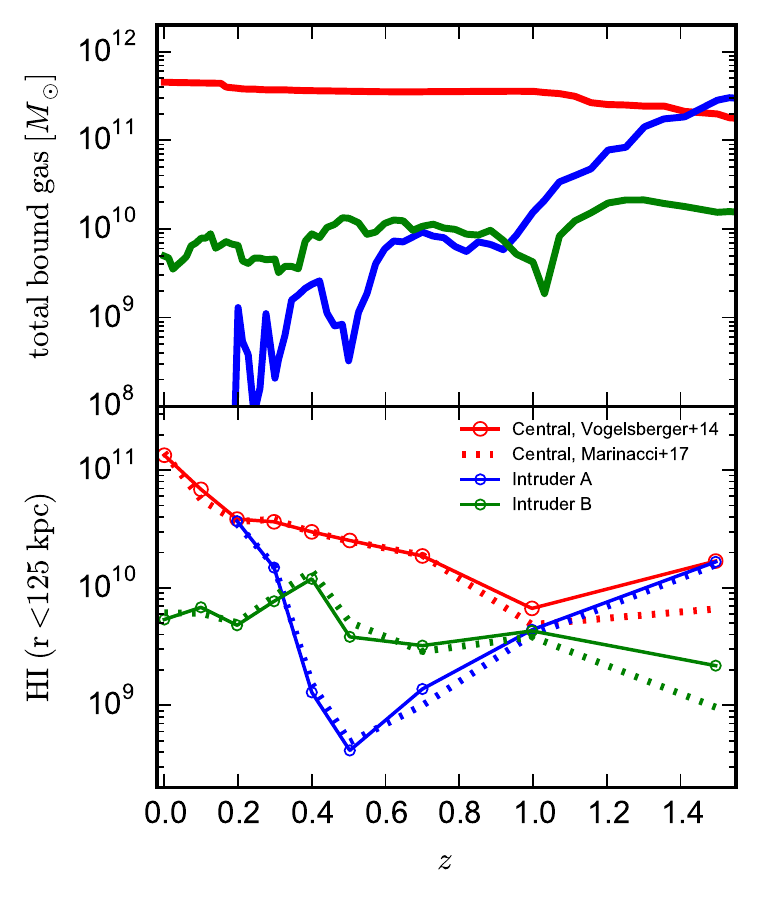}\\
\vspace{-0.25cm}
\caption{\label{fig:gas_mass_evolution}
\textit{Top panel}: Total gravitationally bound gas mass as a function 
of redshift. While Intruder A has much gas until $z = 1$, it loses most of the gas 
since then. On the other hand, Intruder B has retained most of its 
gas mass until $z = 0$, despite having a lower 
$V_{\rm max} = 120\ {\rm km\ s^{-1}}$. 
\textit{Bottom panel}: 
The mass of neutral hydrogen within 125 kpc from the Central and
the two intruders, respectively. For the Central, its neutral hydrogen 
mass has grown by a factor of at least three since $z = 0.2$. 
Among the Central and Intruder A or B, none of the three has
gained this amount of neutral hydrogen prior to the merger event.
The increase of neutral hydrogen within Intruder A since $z = 0.5$ 
is caused by its proximity to the Central because the cold gas is not 
gravitationally bound to Intruder A.}
\end{center}
\end{figure}

All the three players in this cosmic dance are quite massive galaxies. 
In the top panel of Fig.~\ref{fig:gas_mass_evolution}, 
we show the total mass of gravitationally bound gas
as a function of redshift.
While Intruder A was gas-rich at high $z$,
it has lost most of its gas mass since $z = 1$. On the other hand, 
Intruder B has retained most of its gas mass until $z = 0$, despite 
having a lower $V_{\rm max}$. We further compute the mass of
neutral hydrogen within 125 kpc from the three galaxies 
using two methods \citep[][their method 1]{Vogelsberger2014, Marinacci2017}.
The results, shown in the lower panel with solid and dashed lines 
respectively, demonstrate that the two methods are entirely consistent.
The mass of neutral hydrogen within Intruder A has 
gradually declined, so has its total gas mass. The increase in the
amount of neutral hydrogen within Intruder A is due to its proximity 
to the Central, but the cold gas is gravitationally bound only to the 
Central. The Central has grown its neutral hydrogen mass by a 
factor of least three since $z\sim0.2$.

\begin{figure*}
\begin{center}
\begin{tabular}{ccc}
\includegraphics[width=0.32\linewidth]{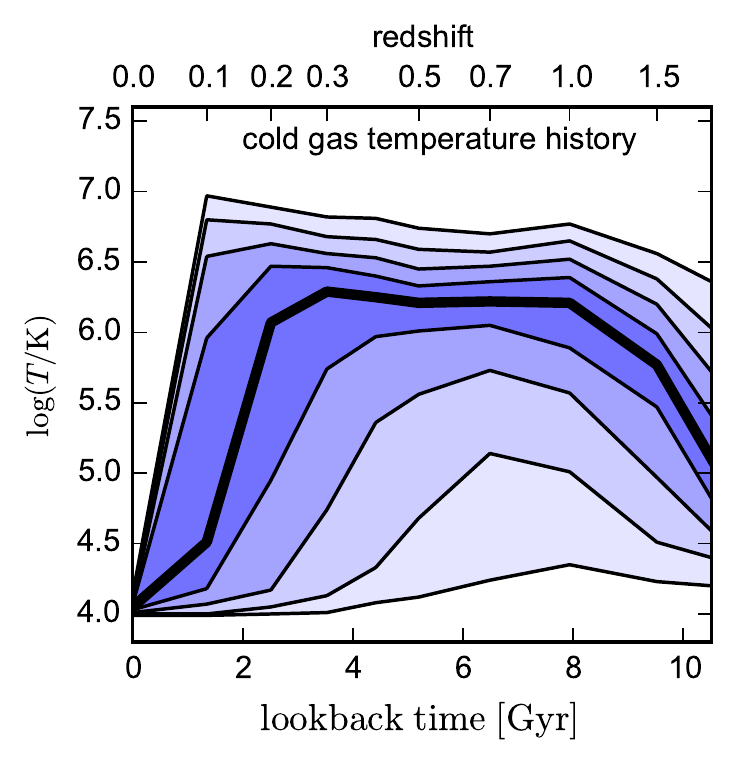} 
\includegraphics[width=0.32\linewidth]{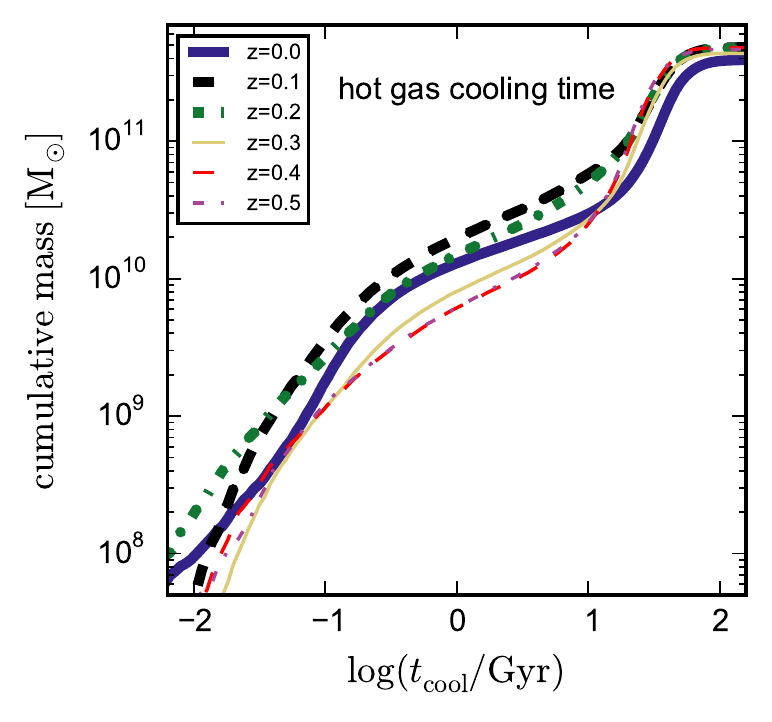} 
\includegraphics[width=0.32\linewidth]{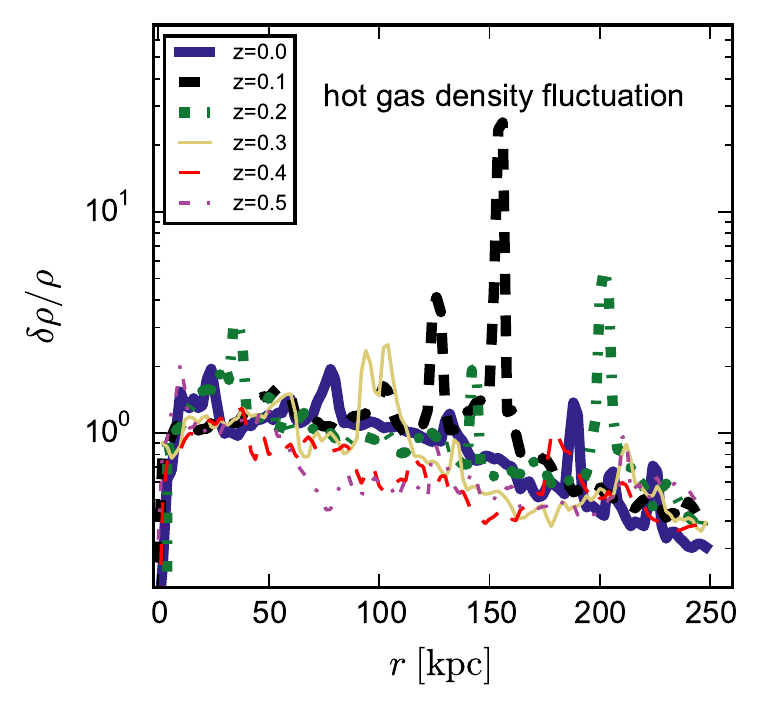}\\
\end{tabular}
\end{center}
\vspace{-0.5cm}
\caption{\label{fig:thermal_history}
\textit{Left panel}: Gas temperature history of the cold gas identified
at $z = 0$ for the Central using tracer particles. We show the 10th, 20th 
up to 90th percentiles at each redshift. The median value of gas temperature
(solid thick line) indicates more than half of the cold gas originated from 
million-degree hot gas at $z = 0.3$. 
\textit{Middle panel}: The distribution of cooling times for the hot halo gas 
within 250 kpc from the Central at each redshift. We observe a strong increase 
of gas mass with cooling time scale shorter than 1 Gyr between $z\sim0.2$ and 0. 
\textit{Right panel}: The density fluctuations $\delta \rho/\rho$ of hot gas, 
a necessary condition for efficient gas cooling, as a function of radius 
at different epochs.}
\end{figure*}

At $z = 0$, the  mass of cold gas within 125 kpc 
from the Central (the size of the final 
gas disk) is $1.8\times10^{11} {\rm M}_{\odot}$,
which is much higher than the total gas mass that can be 
delivered by Intruder A or B. 
The total mass of neutral hydrogen within the Central 
is similarly high, $1.3\times10^{11} {\rm M}_{\odot}$.
To identify the origin of the cold gas, we use 
tracer particles~\citep{Genel2013} to study their thermal histories. 
We select tracer particles within the cold gas disk and tracer particles 
in the hot halo gas, both within 125 kpc from the Central. 

The left panel of Fig.~\ref{fig:thermal_history} shows the 
temperature histories of the cold gas in the Central identified at $z = 0$.
In this plot, we show the gas temperature distribution evolution with redshift.
The evolution of gas temperature shows that 
most of the cold gas mass came from the cooling of million-degree hot
gas, which is triggered at $z\sim0.3$. The central panel shows the distribution 
of cooling times for the hot gas within a radius of 250 kpc. Starting from 
$z\sim0.2$, we observe a strong increase of the gas mass with a cooling time
shorter than 1 Gyr. The right panel shows the density fluctuations as a function 
of radius for hot gas, which is a necessary condition for 
efficient cooling. The largest fluctuations are observed to occur at
$z=0.2$  and $0.1$, when the cold gas mass sharply increased.

\vspace{-0.15in} 
\section{Discussions}
\label{sec:discussions}
The fact that a similarly extended gas/stellar disk as Malin 1 has  
formed in a cosmological simulation is encouraging.
As \cite{Boissier2016} argued, 
the distinct spiral structure and the lack of past bursts of star formation
are difficult to explain by the ring galaxy model \citep{Mapelli2008}. 
A coplanar accretion of a dwarf can lead to an extended 
disk \citep{Penarrubia2006}. However, this process cannot reproduce the 
observed flat gas rotation curve out to 100 kpc \citep{Lelli2010} if Malin 1 
has a similar mass as M31 (as in \citealt{Penarrubia2006}). 
A slowly evolving disk with large angular momentum
and a low level of star formation over cosmic time 
\citep{Boissier2016, Impey1989} is also difficult to realize with
our current cosmology and galaxy formation theory. 
Assuming our Galaxy has a typical spin, a spin parameter twenty 
times larger is exceptional $(5\sigma$ away from the mean) 
although mergers can certainly help build up the spin parameter for 
massive halos \citep{Zjupa2017}. 
More importantly, the disk has yet to survive frequent mergers or 
flybys since high $z$ \citep{RodriguezGomez2017}. 
Overall, it is challenging to reconcile a slowly evolving gas disks 
of $\sim100$ kpc size within a classical inside-out formation 
in $\Lambda$CDM.

In contrast, the mechanism we identified in this study is 
able to explain the extended gas/stellar disks, the 
spiral structure, high gas fraction, as well as a 
flat rotation curve out to 200 kpc. The lack of an age 
gradient across the disk agrees with the recent 
observations by \cite{Boissier2016}. 
The distribution of stellar ages in the extended disk
(continuous star formation since $z=0.3$)
is also consistent with the range (0.1 to 3 Gyr)
found by \cite{Boissier2016}.
The exceptional cold gas mass can 
be explained by the cooling of hot halo gas triggered by an external 
supply of cold gas, i.e. stimulated accretion. 
The entire picture has some similarities to
the one proposed by \cite{Hagen2016}. However, the galaxies in the 
pair are both more massive than typical dwarfs. 
In addition, a large mass fraction of the cold gas originated from
the hot gas halo. 

%Extended and well-defined gas disks are rare in the massive
%galaxies due to AGN feedback in IllustrisTNG \citep[see][]{Torrey2017}. 
We find that only one galaxy contains a massive extended gas 
disk comparable to Malin 1 in the entire volume.
Hot halo gas has been shown to be stable against cooling 
from tiny perturbations \citep{Binney2009, Joung2012}.
On the other hand, favorable conditions for hot halo gas 
to cool are present with an external supply of cold gas 
\citep{Keres2009, Marinacci2011} or turbulence \citep{Gaspari2017, Gaspari2018}.
In particular, the entropy mixing between cold gas and hot gas 
in the turbulent wakes is an effective channel 
for hot gas to cool and condense \citep{Marinacci2011, Armillotta2016}. 
The cooling process in this study can also be viewed as a very special case
in the top-down cold gas condensation framework, 
which plays a crucial role in AGN feeding and star formation in 
early-type galaxies  \citep[for details, see][]{Gaspari2017}.

The large angular momentum of the final gas disk 
is a result of the orbital angular momentum in the cold gas stream. 
We note that the in-spiral of Intruder A has also spun up the hot gas
substantially, from almost no rotation at $z=1$, to a 
specific angular momentum within 250 kpc of 
$1.5\times10^4\, {\rm kpc}\, {\rm km\, s^{-1}}$ at $z = 0.3$, 
which is $\sim$50\% of the value for cold gas. 
The angular momenta of cold and hot gas are also well aligned.
It is possible that the rarity of Malin 1 analogues is a 
consequence of requiring a pair of galaxies on a specific in-spiral orbit 
with much cold gas in at least one member. Model refinements with 
higher resolution would be useful to determine the relative importance 
of each factor at play and reduce the numerical mixing.

Observationally, previous studies have hinted at a possible perturber 
close to Malin 1 \citep{Moore2006, Reshetnikov2010}. Our study 
suggests yet another potential culprit in building up the massive 
gas disk may still be hundreds of kpc away, which could be 
SDSSJ123708.91+142253.2 \citep{Galaz2015}. 
Some neutral hydrogen bridge between SDSSJ123708.91+142253.2 
and Malin 1 will be a strong evidence for the scenario we discussed here.
Also, the presence of an extended old stellar halo around Malin 
1 can be revealed, in principle, with deep photometry to confirm it is
a massive galaxy. Due to the large distance of Malin 1, these observational 
tests are non-trivial. Fortunately, there are similar nearby galaxies such as 
UGC 1382 \citep{Hagen2016} which offer a more feasible alternative. 
Another important test is to establish whether the kinematics and 
metallicity pattern of the outer extended stellar/gas disk are distinct 
from the central regions, as expected from our scenario.

\vspace{-0.15in} 
\section{Conclusions}
\label{sec:conclusions}
We have found a Malin 1 analogue in the 100 Mpc 
IllustrisTNG volume. This galaxy reproduces well the observed 
features of Malin 1's vast disk. The exceptional disk size is a result of hot 
gas cooling triggered by an ample supply of cold gas due to 
a pair of in-falling galaxies. This formation mechanism could reconcile  
the peculiarities of Malin 1's extended stellar disk within current galaxy 
formation theory in $\Lambda$CDM. 

Finally, we note that the conditions that led to the
formation of the simulated galaxy are similar to the present state of the
Milky Way.  Our Galaxy is interacting with a pair of galaxies, the 
Magellanic Clouds, and an extended gas structure, the Magellanic 
Stream.  Theories suggest that the Clouds are on their first 
passage through the Milky Way \citep{Besla2007} and that the 
Stream resulted from the interaction between the Clouds 
\citep{Besla2010}.  Based on these similarities to the object in
IllustrisTNG, we speculate that the Milky Way may
develop an extended disk of its own in the future through the stimulated
accretion of gas from its hot halo.

\vspace{-0.25in}

\section*{ACKNOWLEDGEMENTS}
We thank Lea Hagen and Mark Seibert for stimulating discussions. 
YL acknowledges support from NSF grants AST-0965694, AST-1009867, 
AST-1412719, and MRI-1626251.
M.G. is supported by NASA through Einstein Postdoctoral Fellowship 
Award Number PF5-160137 issued by the Chandra X-ray Observatory 
Center, which is operated by the SAO for and on behalf of NASA under 
contract NAS8-03060. Support for this work was also provided by 
Chandra grant GO7-18121X. 
VS, RP, and RW acknowledge support through the European Research
Council under ERCStG grant EXAGAL-308037.  
VS, RP, RW, DX, and JZ would like to thank the
Klaus Tschira Foundation at HITS. VS and JZ acknowledge support from the 
Deutsche Forschungsgemeinschaft through Transregio 33, `The Dark Universe'. 
The IllustrisTNG flagship simulations were
run on the HazelHen Cray XC40 supercomputer at the High-Performance
Computing Center Stuttgart (HLRS) as part of project GCS-ILLU of the
Gauss Centre for Supercomputing (GCS).  VS also acknowledges support
through sub-project EXAMAG of the Priority Programme 1648 `Software for
Exascale Computing' of the German Science Foundation. MV acknowledges
support through an MIT RSC award, the support of the Alfred P. Sloan
Foundation, and support by NASA ATP grant NNX17AG29G.  
PT acknowledges support from NASA through Hubble Fellowship grant 
HST-HF2-51384.001-A awarded by the STScI, which is operated by 
the Association of Universities for Research in Astronomy, Inc., for NASA, 
under contract NAS5-26555. The Flatiron Institute is supported by the 
Simons Foundation. 

\vspace{-0.15in} 

%Ancillary and test runs of the project were also run on the compute cluster
%operated by HITS, on the Stampede supercomputer at TACC/XSEDE
%(allocation AST140063), at the Hydra and Draco supercomputers at the
%Max Planck Computing and Data Facility, and on the Harvard computing
%facilities supported by FAS.

%%%%%%%%%%%%%%%%%%%%%%%%%%%%%%%%%%%%%%%%%%%%%%%%%%
%%%%%%%%%%%%%%%%%%%%%%%%%%%%%%%%%%%%%%%%%%%%%%%%%%s
%\bibliographystyle{apj}
\bibliographystyle{apj}
\bibliography{ref} 

\iffalse

\fi

% Don't change these lines
%\bsp	% typesetting comment
\label{lastpage}
\end{document}